\documentclass[12pt]{iopart}
\usepackage{iopams}  
\usepackage{graphicx}
\usepackage{cases}

\newcommand{\be}{\begin{equation}}
\newcommand{\ee}{\end{equation}}
\newcommand{\D}{\mathrm{d}}  

\begin{document}

\title{GYOTO: a new general relativistic ray-tracing code}

\author{F H Vincent$^{1,2}$, T Paumard$^{1}$,
E Gourgoulhon$^{2}$ and G Perrin$^{1}$}
\address{$^{1}$ LESIA, Observatoire de Paris, CNRS,  Universit\'e Pierre et Marie Curie, Universit\'e  Paris Diderot, 5 place Jules Janssen, 92190 Meudon, France\\
$^{2}$ LUTH, Observatoire de Paris, CNRS, Universit\'e Paris Diderot, 5 place Jules Janssen, 92190 Meudon, France}
\ead{frederic.vincent@obspm.fr}

\begin{abstract}
	GYOTO, a general relativistic ray-tracing code, is presented. It aims at computing images of astronomical bodies in the vicinity of compact objects, as well as trajectories of massive bodies in relativistic environments.
	This code is capable of integrating the null and timelike geodesic equations not only in the Kerr metric, but also in any metric computed numerically within the 3+1 formalism of general relativity.
	Simulated images and spectra have been computed for a variety of astronomical targets, such as a moving star or a toroidal accretion structure. 
	The underlying code is open source and freely available. It is user-friendly, quickly handled and very modular so that extensions are easy to integrate. Custom analytical metrics and astronomical targets can be implemented in C++ plug-in extensions independent from the main code.
\end{abstract}

\maketitle

\section{Introduction}

The first developments of general relativistic ray-tracing date back to the 70s, with works regarding the appearance of a star orbiting around a Kerr black hole~\cite{cunningham73}, the derivation of an accretion disk's emitted spectrum in terms of a transfer function~\cite{cunningham75}, and the computation of the image of an accretion disk around a Schwarzschild black hole~\cite{luminet79}.

Since this period, many ray-tracing codes appeared in the literature, mainly dedicated to the computation of spectra and particularly of the $6.4$~keV iron line \cite{fabian89,hameury94,fanton97,fuerst04,li05,wu06,dexter09} or to the computation of images of accretion structures \cite{karas92,marck96,fuerst04,dexter09}. Some authors are primarily interested in simulating phenomena of quasi-periodic oscillations in the vicinity of black holes \cite{broderick06,schnittman06}, the trajectory and appearance of stars orbiting around compact objects \cite{levin08,muller09}, the magnetohydrodynamical effects occuring in the surroundings of the black hole \cite{noble07,dolence09}, the collapsar scenario of long gamma-ray bursts and the impact of neutrino pair annihilation~\cite{birkl07,harikae10}, or the polarization of the propagating radiation \cite{broderick04,zamaninasab10,shcherbakov11}.
In this context, one may wonder about the relevance of a new ray-tracing algorithm. There are two main reasons to do so.

Firstly, it is important when using an algorithm to have access to its source code in order to handle it properly, and if needed to develop it according to one's needs. In this perspective, the code presented in this paper, GYOTO (General relativitY Orbit Tracer of Observatoire de Paris), will be available freely (see section~\ref{s:GYOTO_nutshell}). Moreover, as GYOTO is written in C++, it is very easy to add new structures to the existing code, by using the object oriented aspect of the language. Those new structures, representing additional metrics or astronomical objects, can be compiled as plug-in extensions, independently from the Gyoto code itself. Although many ray-tracing codes are used in the literature, they are rarely made public (a recent exception is the \texttt{Geokerr} code \cite{dexter09}; note also
the public release of \texttt{GeodesicViewer} for computing and analyzing individual
geodesics \cite{MulleG10,MulleF11}).

Secondly, as compared with all other existing ray-tracing codes, GYOTO has a specific feature: it is capable of integrating geodesics in non-analytic, numerically computed metrics. This allows GYOTO to handle objects much more diversified than the standard Kerr black holes. In particular, the possibility to handle non-Kerr, numerical metrics allows to test the no-hair theorem of general relativity as investigated by~\cite{psaltis10} in the context of analytic non-Kerr metrics. This kind of study with non-analytic metrics has not been done so far and will be at hand thanks to GYOTO.

Basically, GYOTO consists in launching null geodesics from an observer's screen, that are integrated backward in time to reach an astrophysical object emitting radiation. Once the photon gets inside the emitting object, the equation of radiative transfer is integrated along the computed geodesic in order to determine the value of the emitted specific intensity that will reach the observer.

\section{Ray-tracing in the Kerr metric}

\subsection{Integration of geodesics}

The first goal of GYOTO is to integrate geodesics in the vicinity of rotating black holes. In this section, we will thus consider ray-tracing in the Kerr metric.

\subsubsection{Geodesic equations}

We use the standard Boyer-Lindquist $(x^{\alpha})=(t,r,\theta,\varphi)$ coordinates to express the metric according to~\cite{mtw73}:

\begin{eqnarray}
\D s^{2} &=& g_{\alpha \beta} \, \D x^{\alpha}\D x^{\beta} \nonumber \\ 
			  &=&  -\left(1 - \frac{2 \,M\, r}{\Sigma} \right) \D t^{2}   
- \frac{4 \,M \,a \,r \,\sin^{2}\theta}{\Sigma} \, \D t\, \D \varphi+ \frac{\Sigma}{\Delta}\, \D r^{2} + \Sigma\, \D \theta^{2} \nonumber \\ 
                             & & + \left( r^{2} + a^{2} + \frac{2\, M\, a^{2}\, r\, \sin\theta}{\Sigma} \right) \sin^{2} \theta \, \D \varphi^{2} , 
\end{eqnarray}
where $M$ is the black hole's mass, $a$ its spin parameter ($a M$ being the total angular momentum), $\Sigma \equiv r^{2} + a^{2} \,\cos^{2}\theta$ and $\Delta \equiv r^{2} -2 \,M \,r + a^{2}$.

For a particle of mass $\mu$ and 4-momentum $p_\alpha$, two constants of motion are provided
by the spacetime symmetries: the energy ``at infinity'' $E\equiv-p_{t}$  and the axial component of the angular momentum $L\equiv p_{\varphi}$. The constant nature of $E$ and $L$ results respectively from
the spacetime stationarity and the spacetime axisymmetry. 
In addition, the specific structure of Kerr spacetime gives birth to a third constant of motion, the
Carter constant \cite{carter68}:
\begin{equation}
\label{carter}
Q \equiv p_{\theta}^{2} + \cos^{2} \theta \left[ a^{2}(\mu^{2} - E^{2}) + \sin^{-2}\theta \, L^{2}\right].
\end{equation}
The problem is thus completely integrable by using the four constants $(\mu, E, L,Q)$.



Following~\cite{levin08}, we use a Hamiltonian formulation for the equations of geodesics, which consists in using the variables $(t,r,\theta,\varphi,p_{t},p_{r},p_{\theta},p_{\varphi})$ to express the equations of motion according to
\numparts
\begin{eqnarray}
\label{eqmotion}
\dot{t} &=& \frac{1}{2\, \Delta \Sigma} \frac{\partial}{\partial E} \left( R + \Delta \Theta \right) \\ 
\dot{r} &=& \frac{\Delta}{\Sigma}p_{r} \\ 
\dot{\theta} &=& \frac{1}{\Sigma}p_{\theta} \label{e:dot_theta} \\ 
\dot{\varphi} &=& -\frac{1}{2\, \Delta \Sigma} \frac{\partial}{\partial L} \left( R + \Delta \Theta \right) \\ 
\dot{p_{t}} &=& 0 \label{e:pt}\\ 
\dot{p_{r}} &=& -\left( \frac{\Delta}{2\, \Sigma}Ê\right)' p_{r}^{2} - \left( \frac{1}{2\, \Sigma}\right)' p_{\theta}^{2} + \left( \frac{R+\Delta \Theta}{2 \, \Delta \Sigma} \right)' \\ 
\dot{p_{\theta}} &=& -\left( \frac{\Delta}{2\, \Sigma}Ê\right)^{\theta} p_{r}^{2} - \left( \frac{1}{2\, \Sigma}Ê\right)^{\theta} p_{\theta}^{2} + \left( \frac{R+\Delta \Theta}{2 \, \Delta \Sigma} \right)^{\theta} \\
\dot{p_{\varphi}} &=& 0, \label{e:pph}
\end{eqnarray}
\endnumparts
where the superscripts $'$ and $^{\theta}$ denote differentiation with respect to $r$ and $\theta$ respectively, a dot denoting differentiation with respect to proper time (for timelike geodesics) or to an affine parameter (for null geodesics). In addition, the following abbreviations have been used:
\begin{eqnarray}
\Theta &\equiv &Q - \cos^{2} \left[ a^{2} (\mu^{2}-E^{2}) + \sin^{-2}\theta L^{2}\right] \\
R&\equiv&\left[ E (r^{2} + a^{2}) - a L\right]^{2} - \Delta \left[ \mu^{2} r^{2} + (L- a E)^{2} + Q \right]. 
\end{eqnarray}

\subsubsection{Implementation}

Let us consider the integration of a single null geodesic by GYOTO. The initial conditions are the position of the observer and the direction of incidence of the photon. These quantities allow to determine the tangent vector to the photon's geodesic at the observer's position. 

The integration is then performed backward in time, from the observer towards a distant astrophysical objects. This backward integration is of particular use when the target object has a large angular size on the observer's sky: a large fraction of the computed geodesics will hit the object. If the integration was done from the object, most of the geodesics would not reach the observer, leading to a waste of computing time. However, owing to its modular nature, Gyoto could be fairly easily extended to support the forward ray-tracing paradigm.

The equations of motion are solved by means of a Runge-Kutta algorithm of fourth order (RK4), with an adaptive step (see e.g.~\cite{numrec}). The integration goes on until one of the following stop conditions is fulfilled:

\begin{itemize}
\item the photon reaches the emitting object (however, see section~\ref{seq:radtrans} for the particular treatment inside optically thin objects),
\item the photon escapes too far from the target object (what ``far'' means depends of the kind of object considered),
\item the photon approaches too closely to the event horizon. This limit is typically defined as when the photon's radial coordinate becomes only a few percent larger than the radial coordinate of the event horizon. However this limit must not be too large in order not to stop the integration of photons swirling close to the black hole before reaching the observer, thereby creating higher order images.
\end{itemize}

In order to ensure a proper integration, the conservation of the constants of motion is continuously checked, and the solution of the integration is modified to impose this conservation. This choice of modifying the computed coordinates to ensure the conservation of constants is a consequence of our choice of Hamiltonian formulation for the geodesic equations. The usual form of Kerr geodesic equations (see e.g.~\cite{mtw73}), which directly enforces the conservation of constants, involves square roots that lead to sign indeterminations. The 
Hamiltonian formulation allows to get rid of this problem \cite{levin08}, at the expense of slightly modifying the solution as explained below.

The constancy of $E$ and $L$ is directly enforced by equations~(\ref{e:pt}) and~(\ref{e:pph}). The constancy of $Q$ is imposed by modifying the value of $\dot{\theta}$ given by the RK4 algorithm, according to (\ref{carter}) and (\ref{e:dot_theta}):
\begin{equation}
\dot\theta = \pm \frac{1}{\Sigma} \sqrt{Q - \cos^{2} \theta \left[ a^{2}(\mu^{2} - E^{2}) + \sin^{-2}\theta \, L^{2}\right] } , 
\end{equation}
where the $\pm$ sign is chosen according to the sign of $\dot\theta$ resulting from the 
RK4 algorithm. 
Finally, the constancy of the 4-momentum's norm is ensured by modifying the $\dot{r}$ coordinate furnished by the RK4 algorithm. For instance, when considering a null geodesic, the value of $\dot{r}$ is chosen to enforce the relation 
\begin{equation}
\label{rdotnorm}
g_{tt}\,\dot{t}^{2}+2\,g_{t\varphi}\,\dot{t}\,\dot{\varphi}+g_{rr}\,\dot{r}^{2}+g_{\theta\theta}\,\dot{\theta}^{2}+g_{\varphi\varphi}\,\dot{\varphi}^{2} = 0.
\end{equation}
Thus
\begin{equation}
\label{rdotnormbis}
\dot{r} = \pm \sqrt{-\frac{g_{tt}\,\dot{t}^{2}+2\,g_{t\varphi}\,\dot{t}\,\dot{\varphi}+g_{\theta\theta}\,\dot{\theta}^{2}+g_{\varphi\varphi}\,\dot{\varphi}^{2}}{g_{rr}}}.
\end{equation}
where the $\pm$ sign is chosen according to the sign of $\dot r$ resulting from the 
RK4 algorithm. 
However, in order not to depart too much from the RK4 output, the maximum modification of $\dot{\theta}$ and $\dot{r}$ is limited to $1\%$ of their initial values.

The numerical accuracy of GYOTO is investigated in the Appendix where a convergence test is given.

\subsection{Radiative transfer and spectra computation}

\subsubsection{Radiative transfer}
\label{seq:radtrans}

Once the null geodesic, integrated backward in time from the observer's screen, reaches the emitting object, the equation of radiative transfer must be integrated along the part of the geodesic that lies inside the emitter. In order to perform this computation, two basic quantities have to be known at each integration step: the emission 
coefficient $j_\nu$ 
and the absorption coefficient $\alpha_\nu$ in the frame of a given observer (typically the 
observer comoving with the emitting matter). These coefficients are related to the specific intensity's increment $\D I_\nu$ as along the geodesic according to 
\begin{equation}	
    \D I_{\nu} 	= - \alpha_{\nu}\,I_{\nu}\,\D s \qquad\mbox{and}\qquad \D  I_{\nu} = j_{\nu}\,\D s ,
\end{equation}
where $\D s$ is the element of length as measured by the observer.
The impact of scattering is not taken into account by the code. However, it is possible, if needed, to include in the absorption and  emission coefficients the effect of scattering. Moreover, the modular nature of GYOTO makes it possible to develop a specific scattering treatment that would be plugged into the existing code.

As shown in~\cite{mihalas84}, the relativistic equation of radiative transfer is
\begin{equation}
\label{radtransfGR}
\frac{\D \mathcal{I}}{\D \lambda} = \mathcal{E} 
- \mathcal{A} \, \mathcal{I} ,
\end{equation}
where $\lambda$ is an affine parameter along the considered geodesic, and where the invariant specific intensity $\mathcal{I}$, the invariant emission coefficient 
$\mathcal{E}$
and the invariant absorption coefficient $\mathcal{A}$ have been used:
\begin{equation}
\mathcal{I}\equiv\frac{I_{\nu}}{\nu^{3}}, \qquad \mathcal{E}\equiv\frac{j_{\nu}}{\nu^{2}}, \qquad \mathcal{A} \equiv \nu \, \alpha_{\nu},
\end{equation}
$\nu$ being the frequency of the radiation. These quantities are invariant in the sense that they do not depend on the reference frame in which they are evaluated.

Let us now consider the reference frame comoving with the fluid emitting the radiation. The invariant equation~(\ref{radtransfGR}) becomes in this frame
\begin{equation}
\label{releqbis}
\frac{\D I_{\nu_{\mathrm{em}}}}{\D \lambda} = \nu_{\mathrm{em}} \left[ j_{\nu_{\mathrm{em}}} - \alpha_{\nu_{\mathrm{em}}} \, I_{\nu_{\mathrm{em}}}\right],
\end{equation} 
where $\nu_{\mathrm{em}}$ is the emitted frequency of the radiation.
It is straightforward to obtain the following relation:
\begin{equation}
\label{incr_paff}
\D \lambda\, \nu_{\mathrm{em}} = \D s_{\mathrm{em}} ,
\end{equation}
where $\D s_{\mathrm{em}}$ is the amount of proper length as measured by the emitter. This expression leads to:
\begin{equation}
\label{releqter}
\frac{\D I_{\nu_{\mathrm{em}}}}{\D s_{\mathrm{em}}} = j_{\nu_{\mathrm{em}}} - \alpha_{\nu_{\mathrm{em}}} \, I_{\nu_{\mathrm{em}}}
\end{equation}
where all quantities are computed in the emitter's frame: this is 
the standard form of the equation of radiative transfer in the emitter's frame.

Equation~(\ref{releqter}) can be immediately integrated between some value $s_{0}$ where the specific intensity is vanishing (the ``opposite'' border of the emitting region) and some position $s$:
\begin{equation}
\label{solbis}
I_{\nu} (s) = \int_{s_{0}}^{s} \mathrm{exp}\left( - \int_{s'}^{s} \alpha_{\nu}(s'') \D s''\right)\,j_{\nu}(s')  \D s'.
\end{equation}
Provided that the absorption and emission coefficients are furnished at each position of spacetime, (\ref{solbis}) allows one to compute the integrated specific intensity. This will be illustrated in section~\ref{astrophobj}.

\subsubsection{Spectra computation}

When considering an optically thick object, the computation of a spectrum is quite straightforward. Let us consider the spectrum emitted by a geometrically thin, optically thick accretion disk around a Schwarzschild black hole. Following~\cite{fanton97}, the specific intensity emitted at a given position of the disk is written
\begin{equation}
I_{\nu_{\mathrm{em}}} \propto \delta(\nu_{\mathrm{em}} - \nu_{\mathrm{line}})\,\epsilon(r)
\end{equation}
where $\delta$ is the Dirac distribution and where $\epsilon(r)$ follows a power law with parameter $p$:
\begin{equation}
\epsilon(r) \propto r^{-p}.
\end{equation}
By means of the invariant intensity $\mathcal{I}=I_{\nu} / \nu^{3}$, the specific intensity observed by a distant observer can be related to the emitted specific intensity according to
\begin{equation}
\label{Iobsem}
I_{\nu_{\mathrm{obs}}} = g^{3}\,I_{\nu_{\mathrm{em}}} ,
\end{equation}
where $\nu_{\mathrm{obs}}$ is the observed frequency, and
\begin{equation}
g \equiv \frac{\nu_{\mathrm{obs}}}{\nu_{\mathrm{em}}}. 
\end{equation}

\begin{figure*}
\centering
	\includegraphics[width=6cm,height=6cm]{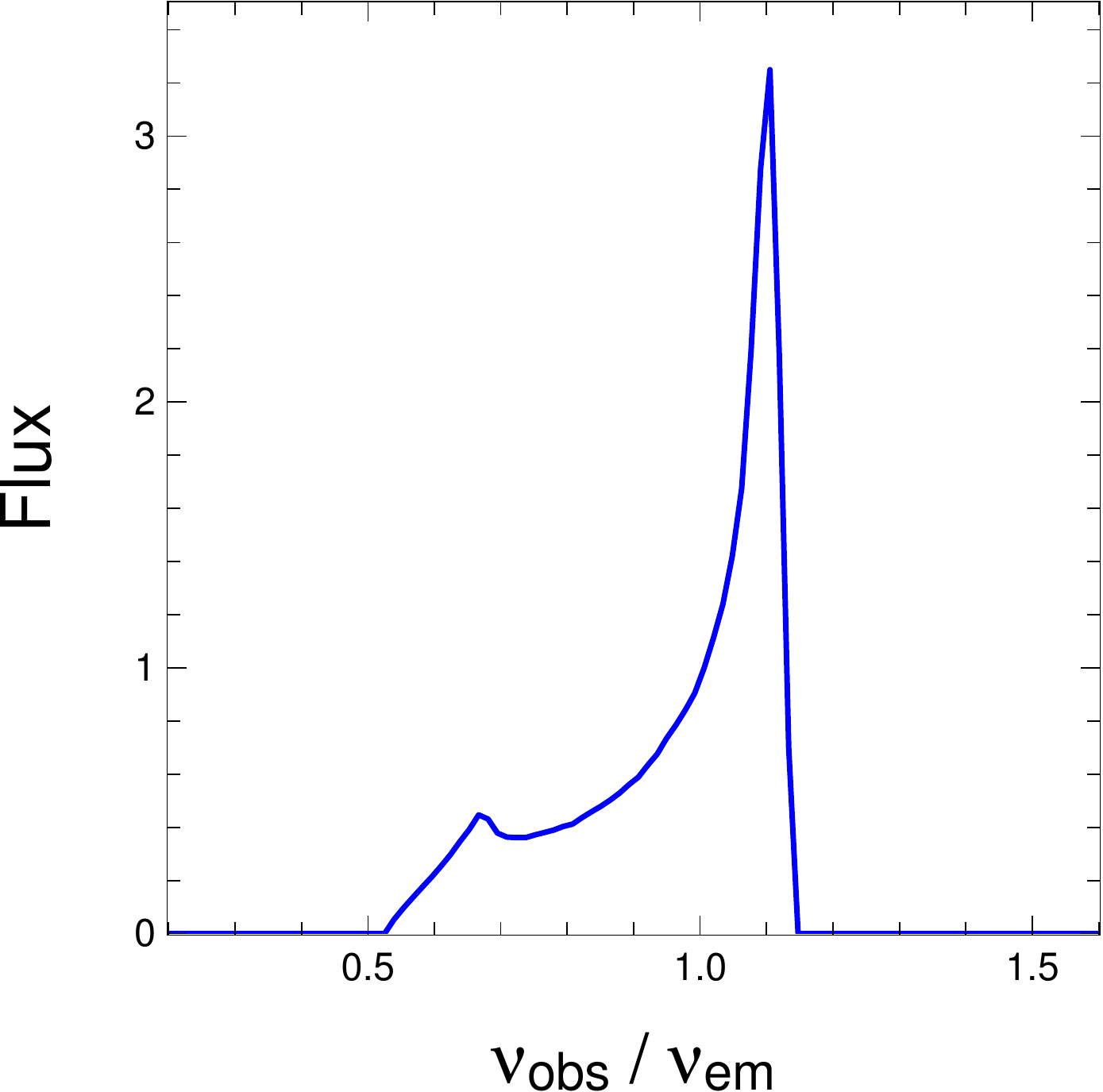}
	\caption{Emission line of a disk around a Schwarzschild black hole of mass $M$ seen under an inclination of $45^{\circ}$. The internal radius of the disk is $r = 6\,M$ and the external radius $r = 10\,M$. The power law parameter is $p=-3$. Only the contribution of the first order image is taken into account. This result can be compared with figure~3 of~\cite{wu06}.}
	\label{imspec}
\end{figure*}

The observed flux $F_{\nu}$ is then related to the observed specific intensity according to
\begin{equation}
\D F_{\nu_{\mathrm{obs}}} = I_{\nu_{\mathrm{obs}}}\,\cos\theta\,\D \Omega ,
\end{equation}
where $\Omega$ is the solid angle under which the emitting element is seen, and $\theta$ is the angle between the normal to the observer's screen and the direction 
of incidence. 
The GYOTO screen is a very simplistic model for a physical telescope equipped which a spectro-imaging camera. It this model, the screen is assumed to be point-like. A pixel on this screen corresponds to a direction on the sky, just like a pixel on a camera detector. Our screen object is defined, among others, by the field-of-view it covers on the sky, the number of samples (or pixels) to cover this field-of-view, and the spectral properties of the detector (number of spectral channels and their wavelength). The screen covers a solid angle $\Omega$ on the sky and each pixel screen corresponds to a small solid angle around a given direction of incidence of the photons. Hence
\begin{equation}
\label{deffluxGyoto}
F_{\nu_{\mathrm{obs}}} = \sum_{\mathrm{pixels}} I_{\nu_{\mathrm{obs}},\mathrm{pixel}}\,\cos(\theta_{\mathrm{pixel}})\,\delta\Omega_{\mathrm{pixel}} ,
\end{equation}
where $I_{\nu_{\mathrm{obs}},\mathrm{pixel}}$ is the specific intensity reaching the given pixel, $\theta_{\mathrm{pixel}}$ is the angle between the normal to the screen and the direction of incidence corresponding to this pixel and $\delta\Omega_{\mathrm{pixel}}$ is the element of solid angle covered by this pixel on the sky, defined as the total solid angle covered by the screen divided by the number of pixels:
\begin{equation}
\delta\Omega_{\mathrm{pixel}} = \frac{2\,\pi\,(1-\cos f)}{N_{\mathrm{pixels}}},
\end{equation}
where $f$ is the angle between the normal to the screen and the most external incident direction of photons on the screen.
Figure~\ref{imspec} shows the resulting emitted spectrum for a Schwarzschild black hole seen under an inclination of $45^{\circ}$.

When considering an optically thin object, the spectrum can also be computed quite easily. It is possible to compute maps of specific intensities for a given panel of $N$ observed frequencies $\nu_{\mathrm{obs},i}$, $1\leq i \leq N$, by using (\ref{solbis}), and to compute the observed flux at these frequencies by using (\ref{deffluxGyoto}). Illustrations of such computations will be given in section~\ref{astrophobj}.

\begin{figure*}
\centering
	\includegraphics[width=5cm,height=5cm]{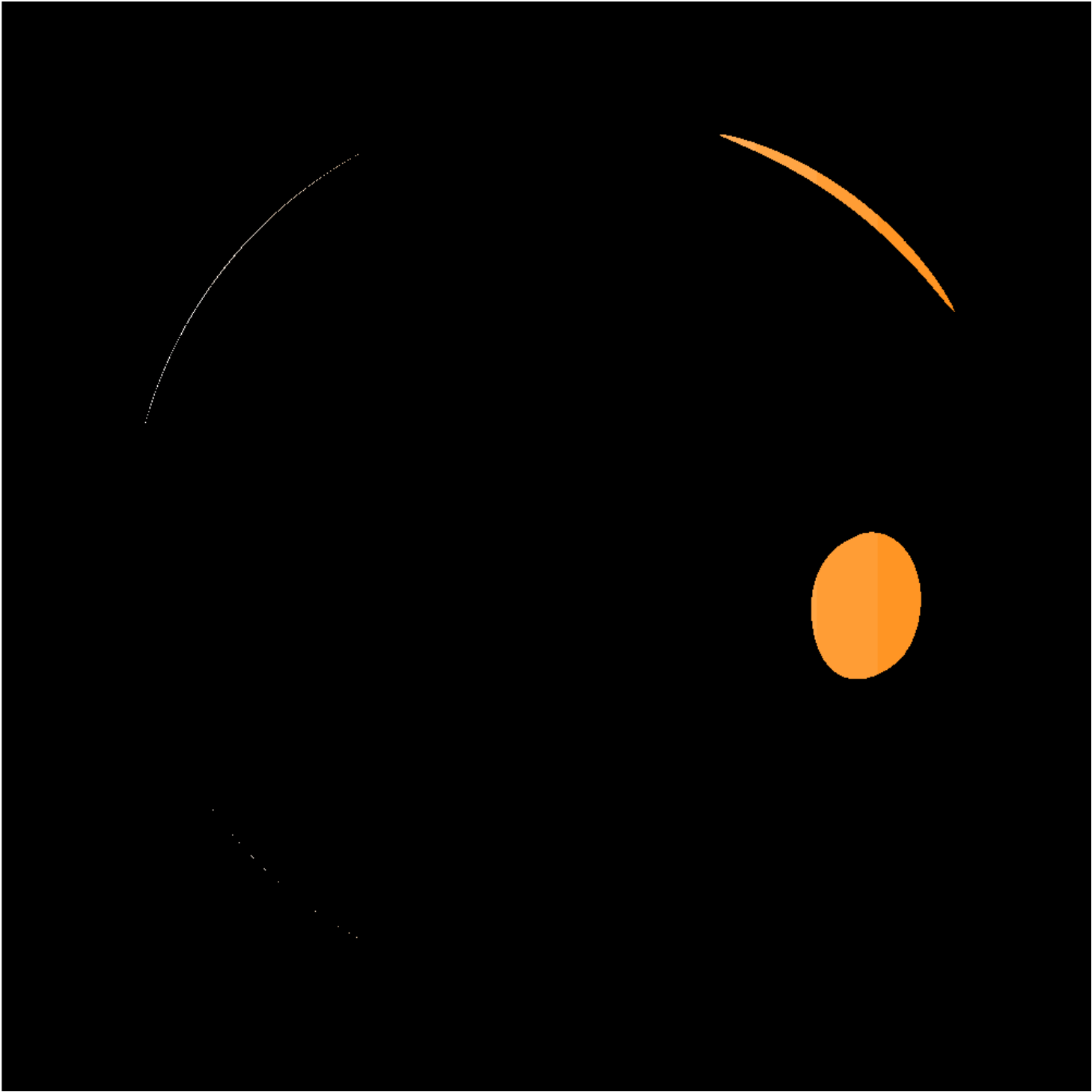}
	\includegraphics[width=9.5cm,height=6.5cm]{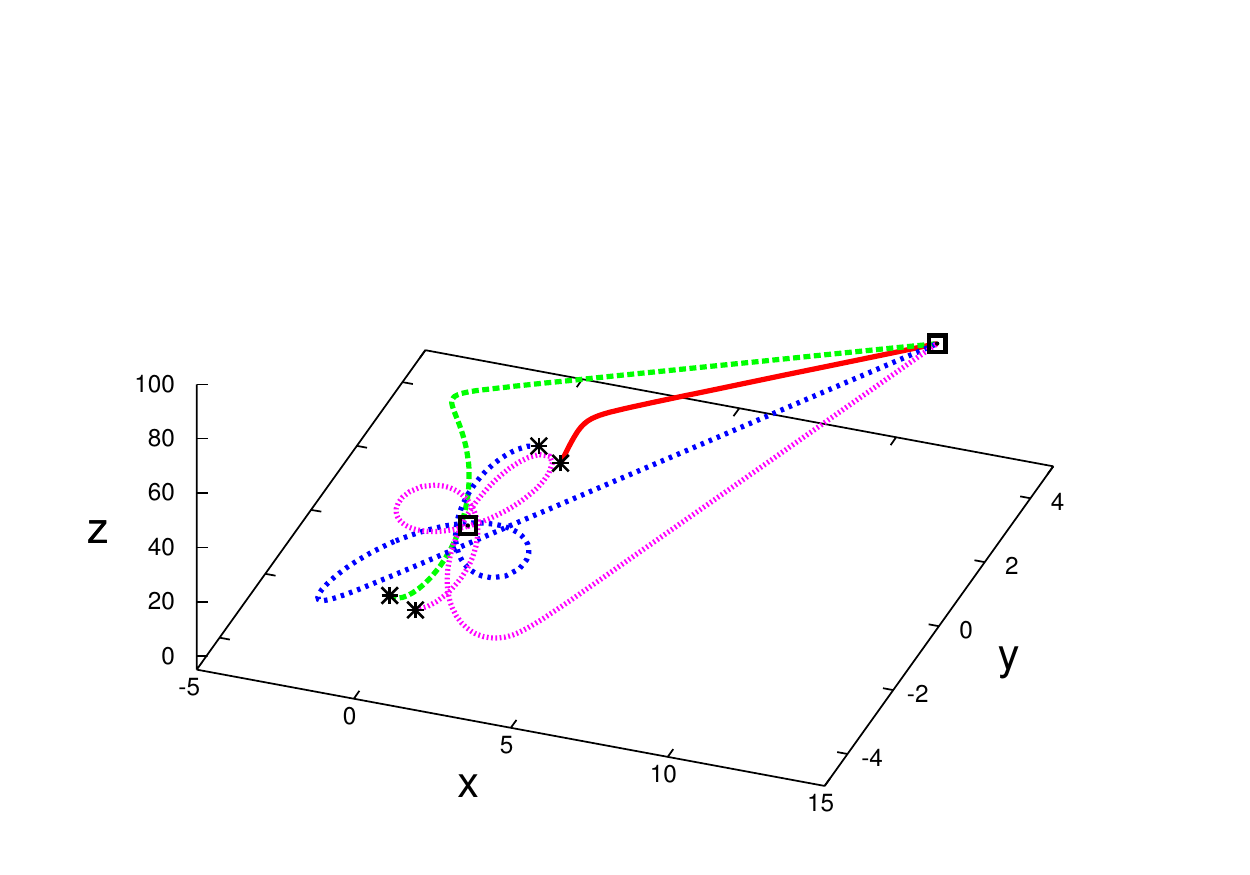}
	\caption{Image of a star at the ISCO of a Kerr black hole with spin parameter
$a=0.9\,M$. 
  \emph{Left:} ray-traced image; the radial coordinate of the observer is $r=100\,M$. \emph{Right:} some specific rays corresponding to the primary (red), secondary (green), tertiary (blue) and quaternary (magenta) images. The various black stars correspond to the source position at the emission of the photons and the black squares depict the black hole and the observer; the axes are labelled in 
 units of $M$. }
	\label{star}
\end{figure*}

\subsection{Implemented astrophysical objects}
\label{astrophobj}

This section describes the various astrophysical objects that are currently 
implemented in GYOTO as potential targets for ray-tracing. 

\subsubsection{Star}

GYOTO can compute the image of a moving star, orbiting around a Kerr black hole. The model for the star is very simple, only the timelike geodesic of the star's center is computed, and the star is defined as the points whose Euclidian distance to the center is less than a given radius $R$, i.e. the points whose Cartesian-like coordinates $(x\equiv r\,\sin \theta\,\cos \varphi,\; y\equiv r\,\sin \theta\,\sin \varphi,\; z\equiv r\, \cos \theta)$ satisfy:
\begin{equation}
(x-x_{c})^{2}+(y-y_{c})^{2}+(z-z_{c})^{2} \leq R^{2},
\end{equation}
where $(x_{c},y_{c},z_{c})$ are the stellar center's coordinates. 
No internal physics nor tidal effects are taken into account. 

Figure~\ref{star} shows the resulting image\footnote{Here, an image is defined as a map of specific intensity $I_{\nu}$.} for a star orbiting on the innermost stable circular orbit (ISCO) of a Kerr black hole of spin parameter $a = 0.9\,M$. Four particular rays are depicted on the right side of the figure, in order to show the trajectory followed by photons responsible for the primary, secondary, tertiary and quaternary images. The corresponding geodesics swirl around the black hole zero, one, two or three times before reaching the object.

The capacity of GYOTO to compute the trajectories of stars around a Kerr black hole will allow to develop relativistic orbit-fitting scripts, in particular by using the Yorick package presented in section~\ref{s:GYOTO_nutshell}.

\begin{figure*}
\centering
	\includegraphics[width=7cm,height=6cm]{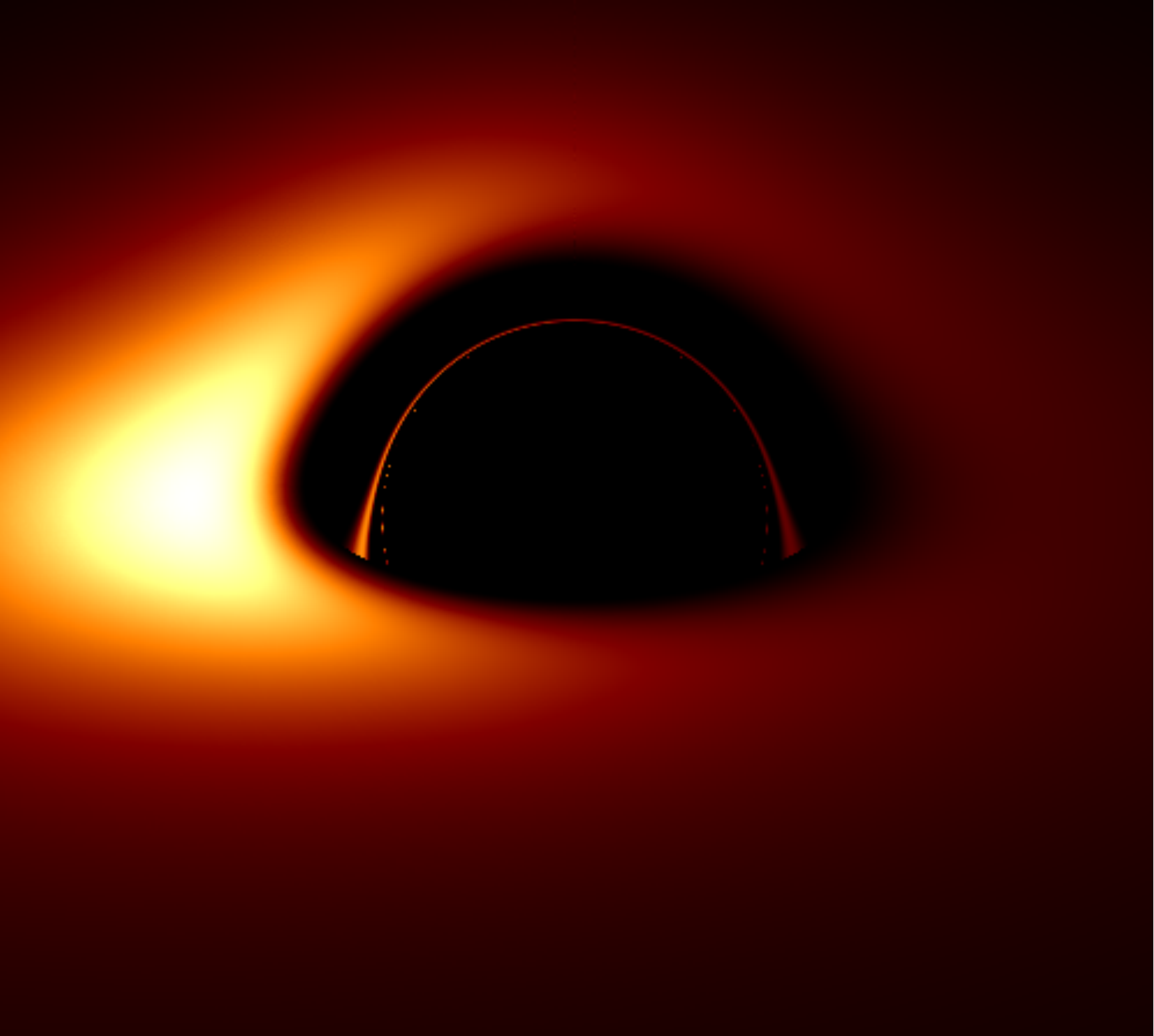}
	\caption{Image of a geometrically thin infinite accretion disk around a Schwarzschild black hole.}
	\label{diskBL}
\end{figure*}

\subsubsection{Thin infinite disk}
\label{sec:thindisk}

A basic object implemented in GYOTO is the geometrically thin infinite accretion disk, with interior radius corresponding to the ISCO \cite{page74}. The expression of the emitted flux is given in~\cite{marck96}, and is proportionnal to the emitted specific intensity provided the emission is isotropic, yielding
\begin{equation}
I_{\nu} \propto \frac{1}{(\rho^{2}-3)\,\rho^{5}}\left\{\rho - \frac{3}{\sqrt{2}}\,\ln\left[(3-2\sqrt{2})\,\frac{\rho+\sqrt{3}}{\rho+\sqrt{3}}\right] \right\} , 
\end{equation}
where $\rho = \sqrt{r}/M$, $M$ being the central black hole's mass.
The resulting image is depicted in 
figure~\ref{diskBL} for the case of a Schwarzschild black hole and an inclination angle
of $70^{\circ}$. It is in good agreement with the results of~\cite{marck96}.

\subsubsection{Rossby wave instability in a thin disk}

\begin{figure*}
  \centering
  \includegraphics[width=7cm,height=7cm]{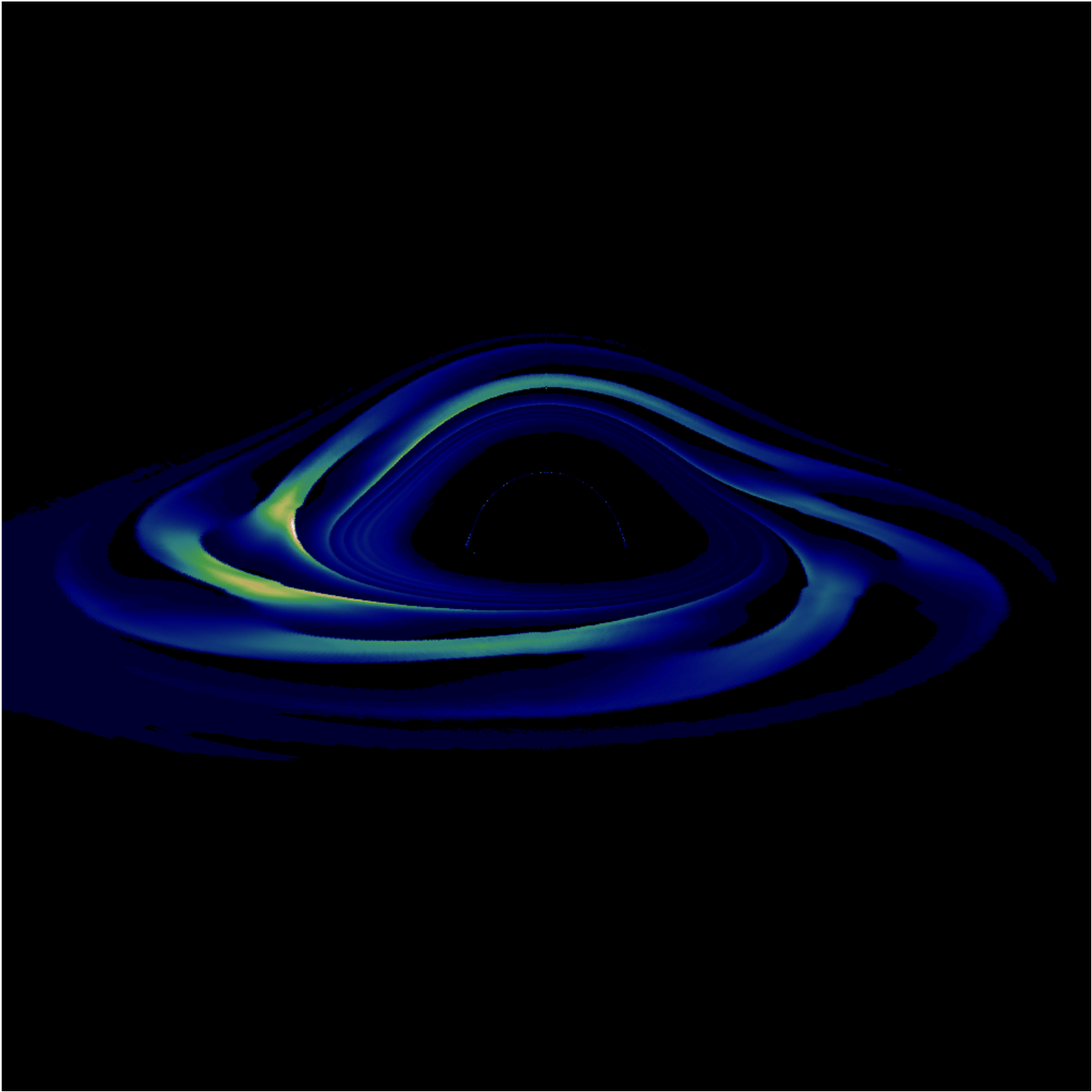}
  \caption{Image of a geometrically thin disk subject to a Rossby wave instability.}
  \label{rossby}
\end{figure*}

In a geometrically thin disk, a Rossby wave instability (RWI) can be triggered if the density profile exhibits an extremum for some value of the radius $r_{\mathrm{RWI}}$
\cite{meheut10}. Density waves will be emitted away from the extremum region, with vortices appearing in the vicinity of $r_{\mathrm{RWI}}$. Such a disk has been implemented in GYOTO, the density profile being computed by means of the VAC code (see~\cite{toth96}). The observed flux emitted by such a disk is computed according to the
prescription in~\cite{falanga07}, as a function of the surface density and radius.
The result is presented in figure~\ref{rossby}, which can be compared with those obtained in~\cite{falanga07}.

\subsubsection{Ion torus}

Ion tori are geometrically thick, optically thin accretion structures with constant 
specific angular momentum. They are the optically thin counterparts of the optically thick Polish doughnuts~\cite{abramowicz78,abramowicz08}.

For a torus made of a perfect fluid, it can be shown \cite{abramowicz78}
that the energy-momentum conservation law
along with the assumption of constant specific angular momentum leads to:
\begin{equation}
\label{cons}
\frac{\nabla_{\mu} p}{p+\epsilon} = \nabla_{\mu} \mathcal{W}
\end{equation}
where $p$ is the pressure, $\epsilon$ is the proper energy density of the fluid and the 
potential $\mathcal{W}$ is related to the covariant component $u_t$ of the 
fluid 4-velocity by $\mathcal{W} = -\ln |u_{t}|$. The isobaric surfaces thus coincide with the equipotential surfaces of $\mathcal{W}$.

The cross-sectional shape of the equipotential surfaces is given in figure~1 of~\cite{abramowicz78}. It appears that one particular surface has a cusp at some $r=r_{\mathrm{crit}}$: it crosses itself in the equatorial plane. Equipotential surfaces contained inside this critical surface are not connected to the central object, thus matter cannot be accreted and swallowed by the black hole. It is thus assumed that the torus physical surface coincides with this critical surface. The central point, $r=r_{\mathrm{central}}$, coincides with the innermost equipotential surface and to the point of maximum pressure. 

It is convenient to introduce the dimensionless parameter
\begin{equation}
\label{w}
w=\frac{\mathcal{W}-\mathcal{W}_{\mathrm{crit}}}{\mathcal{W}_{\mathrm{central}}-\mathcal{W}_{\mathrm{crit}}}.
\end{equation}
The potential (and the pressure) increases continuously between $r_{\mathrm{crit}}$ and $r_{\mathrm{central}}$. Thus, $w$ varies from $0$ (cusp) to $1$ (center) inside the torus. Outside, $w$ can take any other value (positive or negative).

\begin{figure*}
\centering
	\includegraphics[width=7cm,height=5cm]{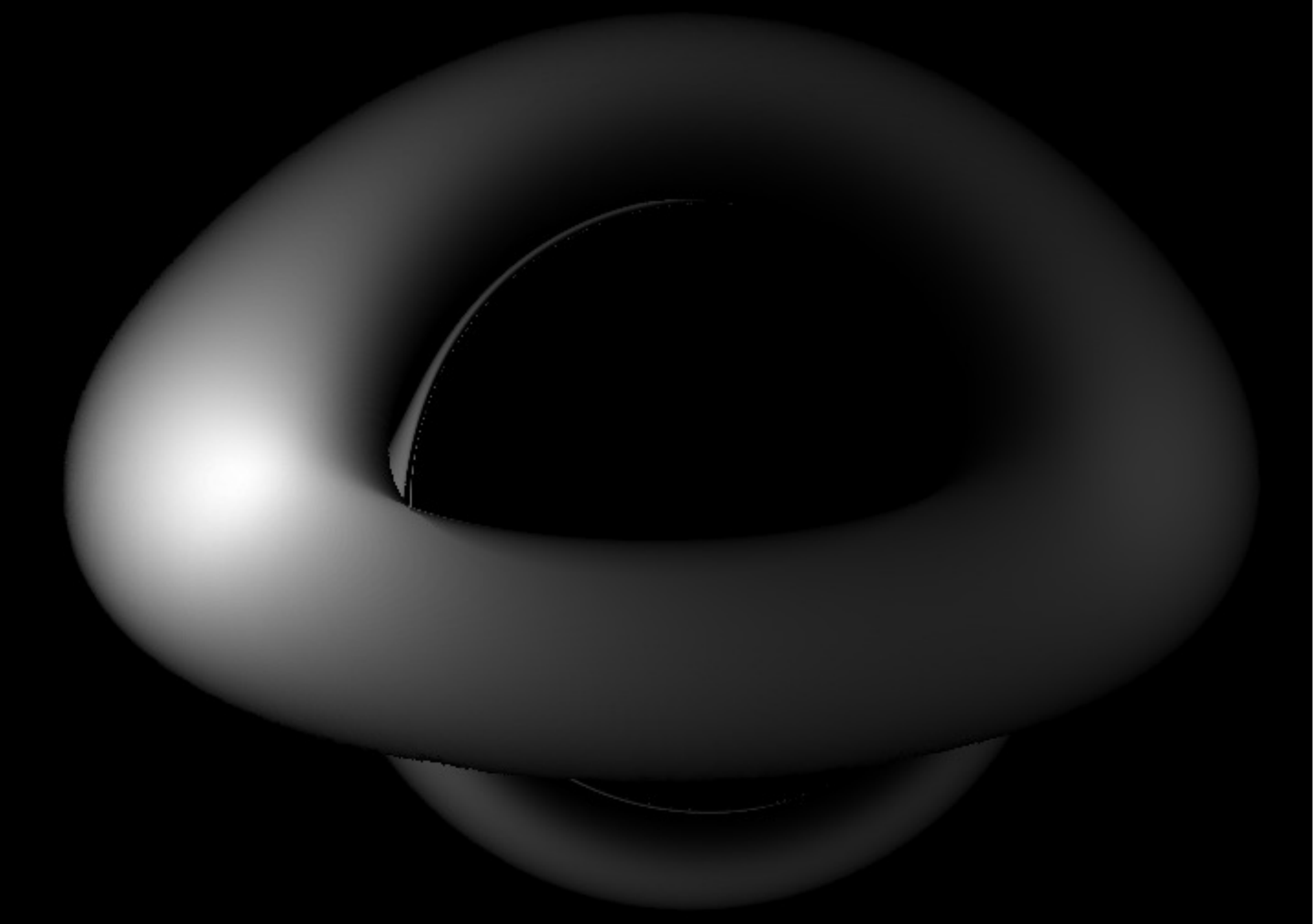}
	\caption{Image of an ion torus around a Kerr black hole of spin parameter $a=0.5\,M$ with  trivial radiative transfer (constant emission coefficient, no absorption).}
	\label{polish-norad}
\end{figure*}
\begin{figure*}
\centering
	\includegraphics[width=7cm,height=7cm]{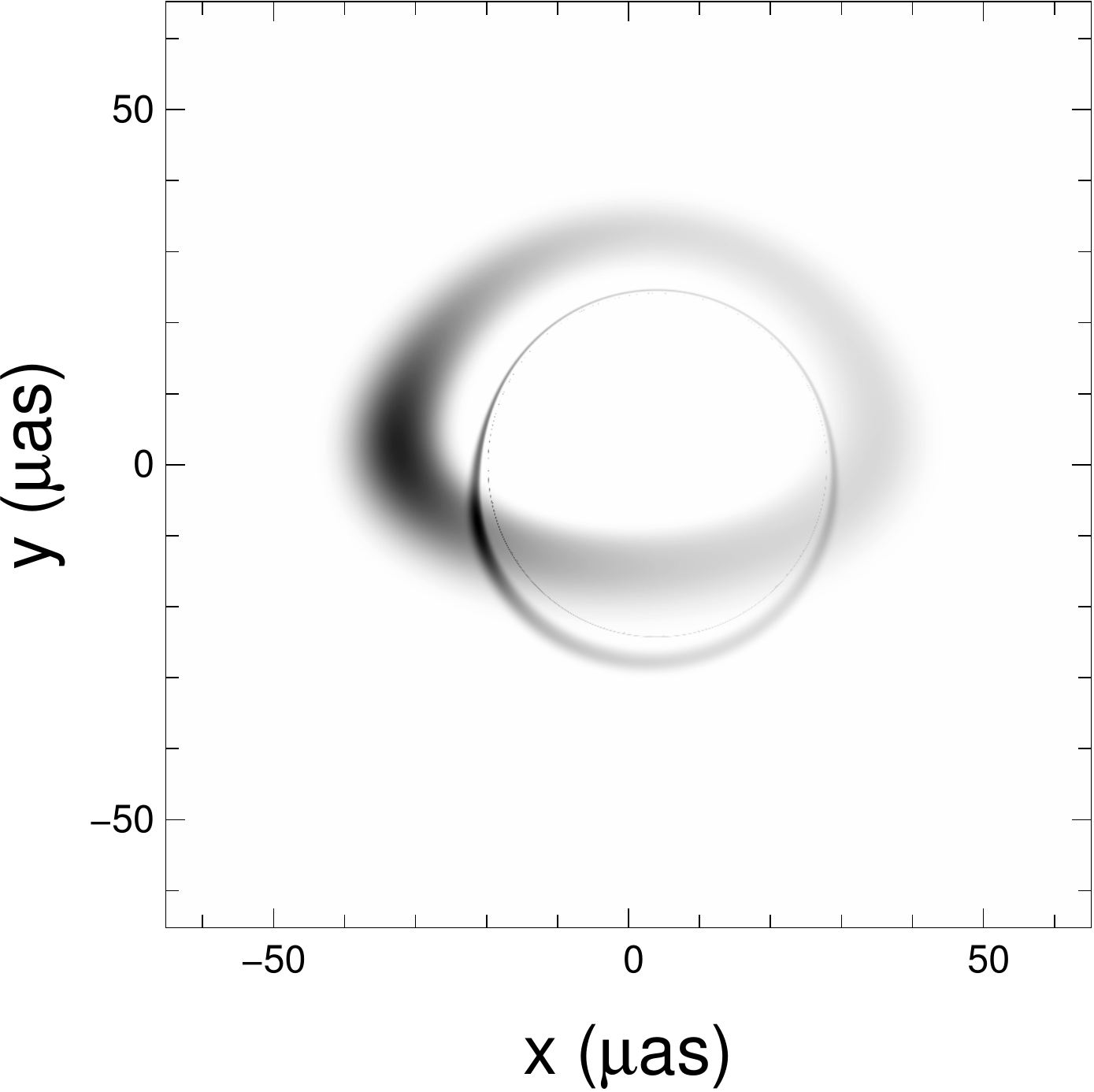}
	\caption{Image of a synchrotron emitting ion torus around a Kerr black hole of mass
$M=4\, 10^6 \, M_\odot$, spin
  parameter $a=0.5\,M$ and placed at the Galactic center, as seen by an observer on Earth. See also~\cite{straub11}.}
	\label{polish}
\end{figure*}

The physics of the radiative transfer used for the ion torus is based on~\cite{narayan95} and is presented in detail in~\cite{straub11}.
Figure~\ref{polish-norad} shows the image of an ion torus around a Kerr black hole with a 
spin parameter $a=0.5\,M$, with trivial radiative transfer included. 
This image can be compared with figure~5 of~\cite{fuerst07}.
Figure~\ref{polish} shows the image of such an ion torus at the Galactic center as seen by an observer on Earth, with radiative transfer including the emission of synchrotron radiation (see~\cite{straub11} for the details of this computation).

\begin{figure*}
  \centering
  \includegraphics[width=7cm,height=7cm]{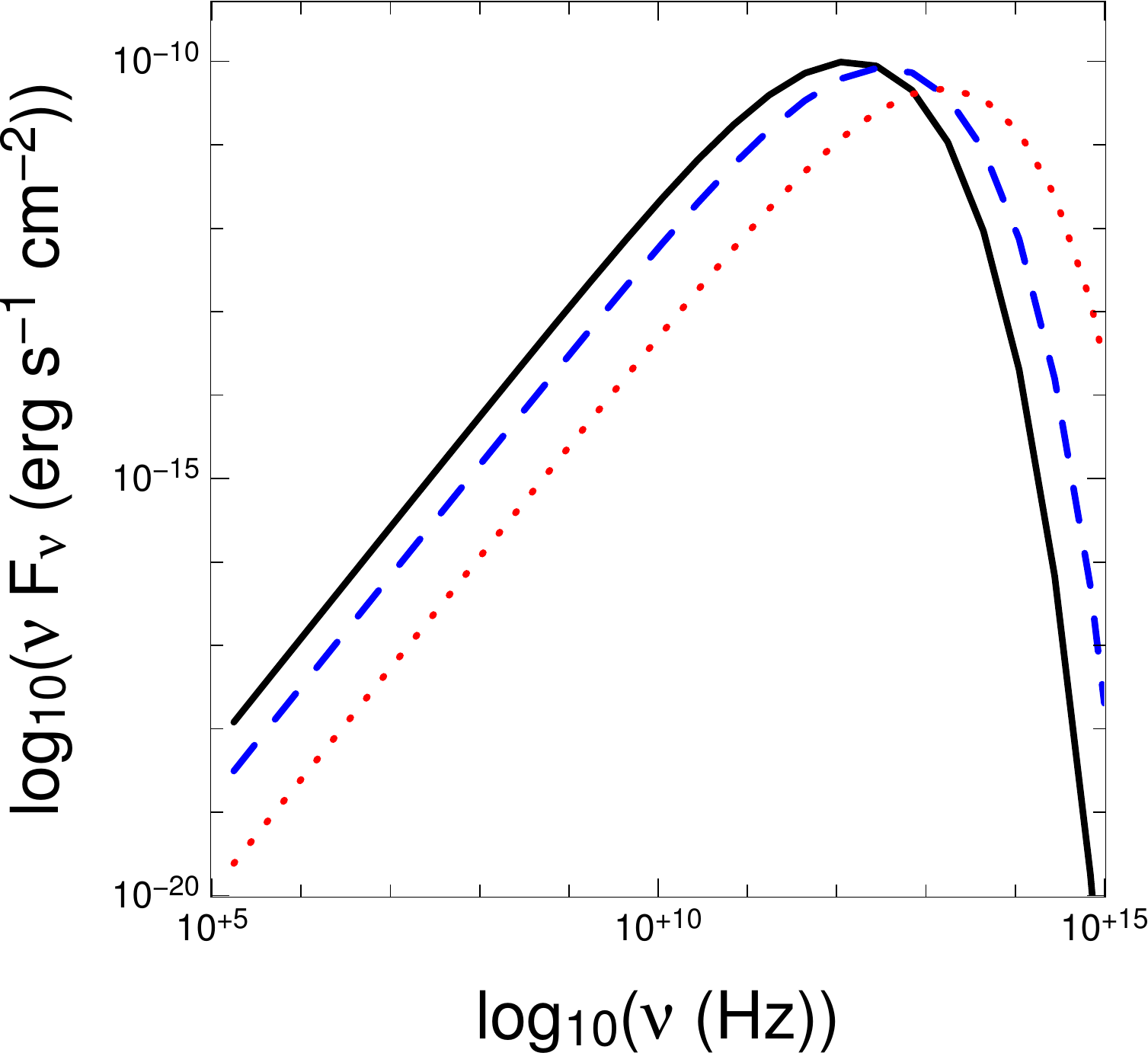}
  \caption{Synchrotron emitted spectrum by the ion torus of figure~\ref{polish} for different spin values. The spin parameter is $a=0$ (solid black), $a=0.5\,M$ (dashed blue) or $a=0.9\,M$ (dotted red). See also~\cite{straub11}.}
  \label{polishspec}
\end{figure*}

Figure~\ref{polishspec} shows the synchrotron emitted spectrum of the previous ion torus. This is an example of the computation of the spectrum of an optically thin object by GYOTO.
Astrophysically interesting perspectives of this kind of computation will be presented in~\cite{straub11}.

\section{Ray-tracing in a numerically computed metric}

\subsection{Introduction}

Many spacetime configurations have been obtained via numerical relativity. 
Most of them have been computed within the 3+1 formalism of general relativity, which relies on the slicing of the 
four-dimensional spacetime by a family of (three-dimensional) hypersurfaces
(see e.g. \cite{gourgoulhon06,Alcub08,BaumgS10}). 
The GYOTO code can perform ray-tracing in such a numerically given metric, 
allowing to consider astrophysical objects beyond the Kerr 
black hole, such as rotating neutron stars, binary neutron stars or black holes, stellar core collapse, etc. 

A spacetime $(\mathcal{M},g_{\alpha\beta})$ is given in a \emph{3+1 form} 
if the following data are available: (i) a family of spacelike hypersurfaces 
$(\Sigma_t)_{t\in\mathbb{R}}$ forming a foliation of $\mathcal{M}$
and (ii) a set $(N,\beta^i,\gamma_{ij},K_{ij})$ of 
fields\footnote{Following standard conventions, Latin indices span $\{1,2,3\}$, whereas Greek ones span $\{0,1,2,3\}$.} on each hypersurface 
$\Sigma_t$ such that $N$ is a strictly positive scalar field, called the 
\emph{lapse function}, $\beta^i$ is a vector field on $\Sigma_t$, called 
the \emph{shift vector}, $\gamma_{ij}$ is a positive definite metric 
on $\Sigma_t$, called the \emph{3-metric} or \emph{induced metric} and 
$K_{ij}$ is a symmetric tensor field on $\Sigma_t$, called the 
\emph{extrinsic curvature tensor} (see e.g. \cite{gourgoulhon06,Alcub08,BaumgS10}). 
If $(x^i)$ are regular coordinates on each $\Sigma_t$, varying smoothly 
from one hypersurface to the next one, then $(x^\alpha) = (t,x^i)$ is a valid
coordinate system for the whole spacetime $(\mathcal{M},g_{\alpha\beta})$. 
From the knowledge of $(N,\beta^i,\gamma_{ij})$, one can reconstruct the 
spacetime metric $g_{\alpha\beta}$ according to 
\begin{equation} \label{e:g_3p1}
  g_{\alpha\beta} \, \D x^\alpha \, \D x^\beta = 
  - N^2 \D t^2 + \gamma_{ij} (\D x^i + \beta^i \D t)
		(\D x^j + \beta^j \D t) . 
\end{equation}

\subsection{GYOTO treatment of 3+1 metrics} \label{s:Gyoto_3p1}

Assuming that the metric is provided in a 3+1 form, two techniques of geodesics integration have been implemented in GYOTO. 

\subsubsection*{Four-dimensional method:} 
From the 3+1 data $(N,\beta^i,\gamma_{ij})$, reconstruct the 4-metric 
$g_{\alpha\beta}$ according to (\ref{e:g_3p1}); then compute the 
Christoffel symbols ${}^4 \Gamma^\alpha_{\ \, \mu\nu}$ of $g_{\alpha\beta}$
with respect to the coordinates $(x^\alpha)$ and integrate the geodesic 
equation in the standard 4-dimensional form: 
\begin{equation} \label{e:geod_4D}
	\frac{\D^2 X^\alpha}{\D \lambda^2} + {}^4 \Gamma^\alpha_{\ \, \mu\nu} \frac{\D X^\mu}{\D\lambda}
	\frac{\D X^\nu}{\D\lambda} = 0 , 
\end{equation} 
where $\lambda$ is either the proper time (timelike geodesics) or some affine parameter (null geodesics), the geodesic being defined by 
$x^\alpha = X^\alpha(\lambda)$.  

\subsubsection*{Three-dimensional method:} 
The geodesic is searched in the form 
\be
  x^i = X^i(t) , 
\ee
i.e. the geodesic is parametrized by the coordinate time $t$ and not by $\lambda$. 
In \cite{vincent11}, we have derived a system of differential equations
for the functions $X^i(t)$, which involve only the 3+1 quantities 
$(N,\beta^i,\gamma_{ij},K_{ij})$ and their spatial derivatives 
(denoted by $\partial_i \equiv  \partial/\partial x^i$) : 
\begin{subnumcases}{\label{e:systXV}}
  \frac{\D X^i}{\D t} = & $ \!\!\!\!\!\! N V^i - \beta^i$ \label{e:dXdt} \\[1ex]
  \frac{\D V^i}{\D t} = &   $\!\!\!\!\!\! N\left[ V^i\left( V^j\partial_j\ln N - K_{jk} V^j V^k \right)
  + 2 K^i_{\ \, j} V^j - {}^3\Gamma^i_{\  jk} V^j V^k \right]$ \nonumber\\
  &   $\!\!\!\!\!\!- \gamma^{ij}\partial_j N - V^j \partial_j \beta^i$ . \label{e:dVdt}
\end{subnumcases}
Here the ${}^3\Gamma^i_{\  jk}$ are the Christoffel symbols of the 3-metric
$\gamma_{ij}$ with respect to the coordinates $(x^i)$. The auxiliary variable
$V^i(t) = N^{-1} (\D X^i / \D t + \beta^i)$ is the linear momentum of
the considered particle divided by its energy, both quantities being 
measured by the Eulerian observer (i.e. the observer whose worldline is normal to the $\Sigma_t$ hypersurfaces) \cite{vincent11}.

\subsection{Discretization} \label{s:discretization}

The differential equations (\ref{e:geod_4D}) or (\ref{e:systXV}) are 
integrated by means of a fourth-order Runge-Kutta algorithm. 
This requires the values of the 3+1 fields $(N,\beta^i,\gamma_{ij},K_{ij})$
at arbitrary points. 
If the numerical spacetime has been computed by means of a spectral method
\cite{GrandN09}, this does not cause any great difficulty since by essence
such a method deals with fields and not with values at grid points (see below). 
If the numerical spacetime arises instead from a finite difference method
\cite{Alcub08,BaumgS10}, the fields $(N,\beta^i,\gamma_{ij},K_{ij})$ are known only at the
points of the grid used for the computation. 
One should then devise some 
interpolation procedure to get
the values of the fields $(N,\beta^i,\gamma_{ij},K_{ij})$ and their
derivatives at the points required by the 
Runge-Kutta algorithm. 

Currently the geodesic integration is implemented in GYOTO only for numerical metrics computed by means of spectral methods. 
We are using spherical-type coordinates $(x^i)=(r,\theta,\varphi)$ on $\Sigma_t$
and divide the range of $r$ in $n$ computational domains ($n\geq 2$) : 
$r\in[0,r_0]$, $r\in(r_0,r_1]$, ..., $r\in(r_{n-2},+\infty)$. 
The \emph{spectral method} consists in choosing finite numbers $N_r$, $N_\theta$ and
$N_\varphi$ of degrees of freedom in $r$, $\theta$ and $\varphi$ respectively
(typically $N_{r,\theta,\varphi} \sim 20$) and in expanding 
any scalar field $f$ onto a polynomial basis: 
\be \label{e:spectral_expan}
  f(t,r,\theta,\varphi) = \sum_{k=0}^{N_\varphi-1} \sum_{j=0}^{N_\theta-1} \sum_{i=0}^{N_r-1}  
    \hat f_{ijk}(t) \, R_{ij}(\xi) \, \Theta_{jk}(\theta) \, \Phi_k(\varphi),
\ee
where $\xi\in[0,1]$ ($r\leq r_0$) or $\in[-1,1]$ ($r> r_0$) and is related to $r$ by
\be \label{e:def_xi}
  \xi \equiv  \cases{
    r / r_0  &for $r\leq r_0$ \\
    \frac{2r-r_\ell -r_{\ell-1}}{r_\ell - r_{\ell-1}} &
    for $r_{\ell-1} < r \leq r_\ell$, \quad $\ell\in[1,n-2]$ \\
    1 - \frac{2 r_{n-2}}{r} &for $r > r_{n-2}$
  }
\ee
and $\Phi_k$, $\Theta_{jk}$ and $R_{ij}$ are (trigonometric) polynomial functions.
Working with the \textsc{Lorene} library \cite{lorene}, we choose
\be \label{e:Phi_k}
  \Phi_k(\varphi) \equiv  \cases{
    \cos(m_k\varphi) &for $k$ even \\
    \sin(m_k\varphi) &for $k$ odd, 
  }
\ee
\be
  m_k \equiv  [k/2] \quad\mbox{(integer part of $k/2$)},
\ee
\be \label{e:Theta_jk}
  \Theta_{jk}(\theta) \equiv  \cases{
    \cos(j\theta) & for $m_k$ even \\
    \sin(j\theta) & for $m_k$ odd, }
\ee
\be \label{e:R_i}
  R_{ij}(\xi) \equiv  \cases{
  T_{2i}(\xi) & for $j$ even and $r \leq r_0$\\
  T_{2i+1}(\xi) & for $j$ odd and $r \leq r_0$\\
  T_i(\xi)& for $r> r_0$,} 
\ee
$T_i$ being the Chebyshev polynomial of order $i$, defined by the identity $\cos(i x) = T_i(\cos x)$.
Note that the $\varphi$-basis defined by (\ref{e:Phi_k}) is nothing but a Fourier basis, the field $f$ being obviously periodic in $\varphi$. 
The choice (\ref{e:Theta_jk}) for the $\theta$-basis is motivated by regularity
conditions associated with spherical coordinates (consider for example
$x=r\sin\theta\cos\varphi$ ($m=1$) and $z=r\cos\theta$ ($m=0$)).  
The choice (\ref{e:R_i}) for the $\xi$-basis is motivated by the good properties of the Chebyshev polynomials with respect to the approximation of 
functions \cite{GrandN09}. An alternative choice could have been Legendre polynomials. In (\ref{e:R_i}), the well defined parity of the polynomial 
for $r\leq r_0$ follows from the regularity conditions associated with spherical coordinates around $r=0$. 
Finally note that thanks to the choice (\ref{e:def_xi}) for $\xi$, the last domain 
extends to $r=+\infty$, which is reached for a finite value of $\xi$,
namely $\xi=1$. 

In view of (\ref{e:spectral_expan}), it is clear that within the spectral method, the 
scalar field $f$ is entirely described by the $N_r\times N_\theta\times N_\varphi$
coefficients $\hat f_{ijk}(t)$. For instance, the radial derivative of 
$f$ is evaluated as
\be
  \frac{\partial f}{\partial r}(t,r,\theta,\varphi) = 
  \frac{\D \xi}{\D r} \, \sum_{k=0}^{N_\varphi-1} \sum_{j=0}^{N_\theta-1} \sum_{i=0}^{N_r-1}  
    \hat f_{ijk}(t) \, R'_{ij}(\xi) \, \Theta_{jk}(\theta) \, \Phi_k(\varphi),
\ee
where the derivative $R'_{ij}(\xi)$ of the polynomial $R_{ij}(\xi)$ can be expanded
as 
\be
  R'_{ij}(\xi) = \sum_{p=0}^{N_r-1} a_{ijp} \,  R_{p(j+1)} (\xi) . 
\ee
The index $(j+1)$ in the right-hand side reflects the change of parity induced by the derivative and $a_{ijp}$ are known coefficients which depend on the actual family of polynomials. 

In the course of the Runge-Kunta procedure for the geodesic integration, 
when the value of one of the 3+1 fields is required at a given point, one uses
formula (\ref{e:spectral_expan}) to evaluate it. If the numerical spacetime is 
stationary, the coefficients $\hat f_{ijk}(t)$ do not depend upon $t$ and
are part of the known data. In a dynamical spacetime, one has to evaluate $\hat f_{ijk}(t)$
from the known data, which are the values $\hat f_{ijk}(t_J)$ 
of the coefficients at a series $t_J$ of discrete times --- the coordinate times
at which the numerical spacetime has been computed. To this aim, an interpolation at order 3 is used, using the 4 neighbouring $t_{J}$ to compute the field value at $t$, by means of Neville's algorithm (see e.g.~\cite{numrec}).

Once the coefficients $\hat f_{ijk}(t)$ are known, formula (\ref{e:spectral_expan}) provides a very accurate value of $f$, with an error decaying exponentially
with the numbers of degrees of freedom $N_r$, $N_\theta$ and $N_\varphi$ \cite{GrandN09}. 
As a result, a number of degrees of freedom of order $20$ is generally sufficient to reach
the maximum relative accuracy
allowed in double-precision computing ($\sim 10^{-14}$). Note that formula (\ref{e:spectral_expan}) involves a number of arithmetic operations
which is of the order of $N_r\times N_\theta\times N_\varphi$. 
Consequently the ray-tracing in numerical metrics is far more time consuming than
that in analytical ones. 

\begin{figure*}
\centering
	\includegraphics[width=10cm,height=8cm]{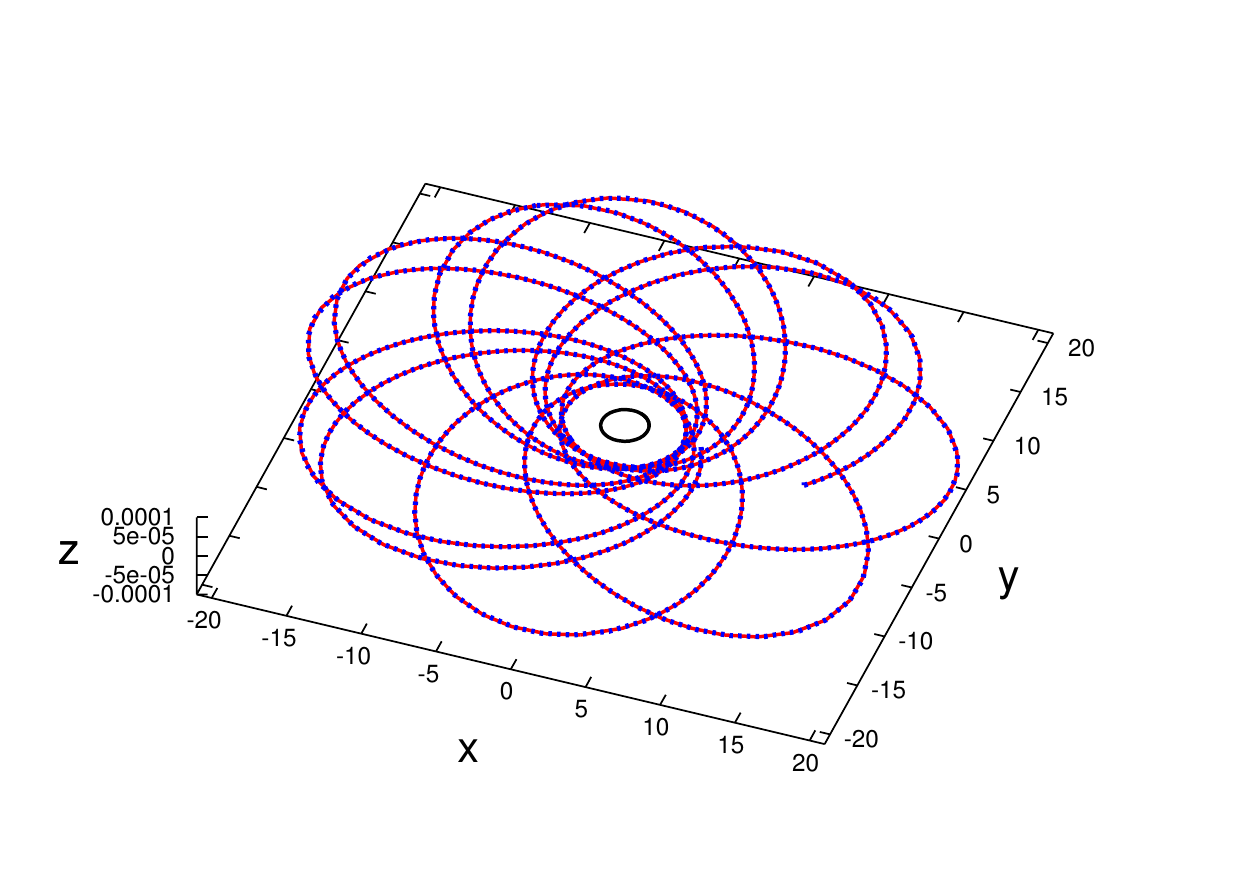}
	\caption{Orbit of a massive particle around a rotating relativistic star, obtained as a timelike geodesic in a numerical metric resulting from the \texttt{Lorene/nrotstar} spectral code. The computation has been performed either by integrating the
4-dimensional geodesic equation (\ref{e:geod_4D}) (solid red line) or by
integrating the 3+1 geodesic system (\ref{e:systXV}) (dotted blue line). The black circle is the central star's equatorial circumference. The axes are graduated in units of 10 km. The central star
has a gravitational mass of $2.1\,M_{\odot}$, is rotating at a frequency of 700~Hz, its equatorial coordinate radius is $15.2$~km and the ratio of the polar to equatorial radii is $0.76$. The orbiting massive particle is restricted to the equatorial plane, with its $Z$ coordinate limited to 
$|Z|< 1\;\mathrm{m}$. The radial coordinate of the star oscillates between a minimum value of around $38$~km and a maximum of $210$~km.}
	\label{test3_1part}
\end{figure*}

\subsection{An illustrative example}

To illustrate the computation of geodesics in a numerical metric, we consider 
a test-mass particle orbiting around a relativistic rotating fluid star. 
Contrary to the black hole case, the metric around the star is not known analytically (except when the star is not rotating, where it reduces to Schwarzschild metric). It has to be computed numerically. For this purpose, we have used the \texttt{Lorene/nrotstar} code \cite{lorene}, which is based
on a spectral method in spherical coordinates (as described in 
section~\ref{s:discretization}) and solves the Einstein equations
according to the framework presented in~\cite{gourgoulhon10}.
The computation of the timelike geodesic is performed via both methods 
exposed in section~\ref{s:Gyoto_3p1}. The results are compared in 
figure~\ref{test3_1part}, revealing a very good agreement between the two methods.

\section{GYOTO code in a nutshell}
\label{s:GYOTO_nutshell}

GYOTO is documented in-depth on its homepage
\texttt{http://gyoto.obspm.fr}. 
Here we simply summarize its most salient features.

Several interfaces are provided to use the GYOTO code:
\begin{itemize}
\item a command line utility (\texttt{gyoto}) allows ray-tracing a single frame. The scenery is specified through a XML\footnote{http://www.w3.org/XML/} file and the output frame is stored in the astronomy standard FITS\footnote{http://fits.gsfc.nasa.gov/} file format;
\item an extension package (\texttt{ygyoto}) for the Yorick\footnote{http://yorick.sourceforge.net/} interpreter allows using GYOTO interactively from a command prompt and writing complex scripts. A graphical user interface (\texttt{gyotoy}), based on this Yorick package, allows to interactively trace a single timelike geodesic in Kerr metric (see figure~\ref{gyotoy});
\begin{figure*}
\centering
	\includegraphics[width=8cm,height=6cm]{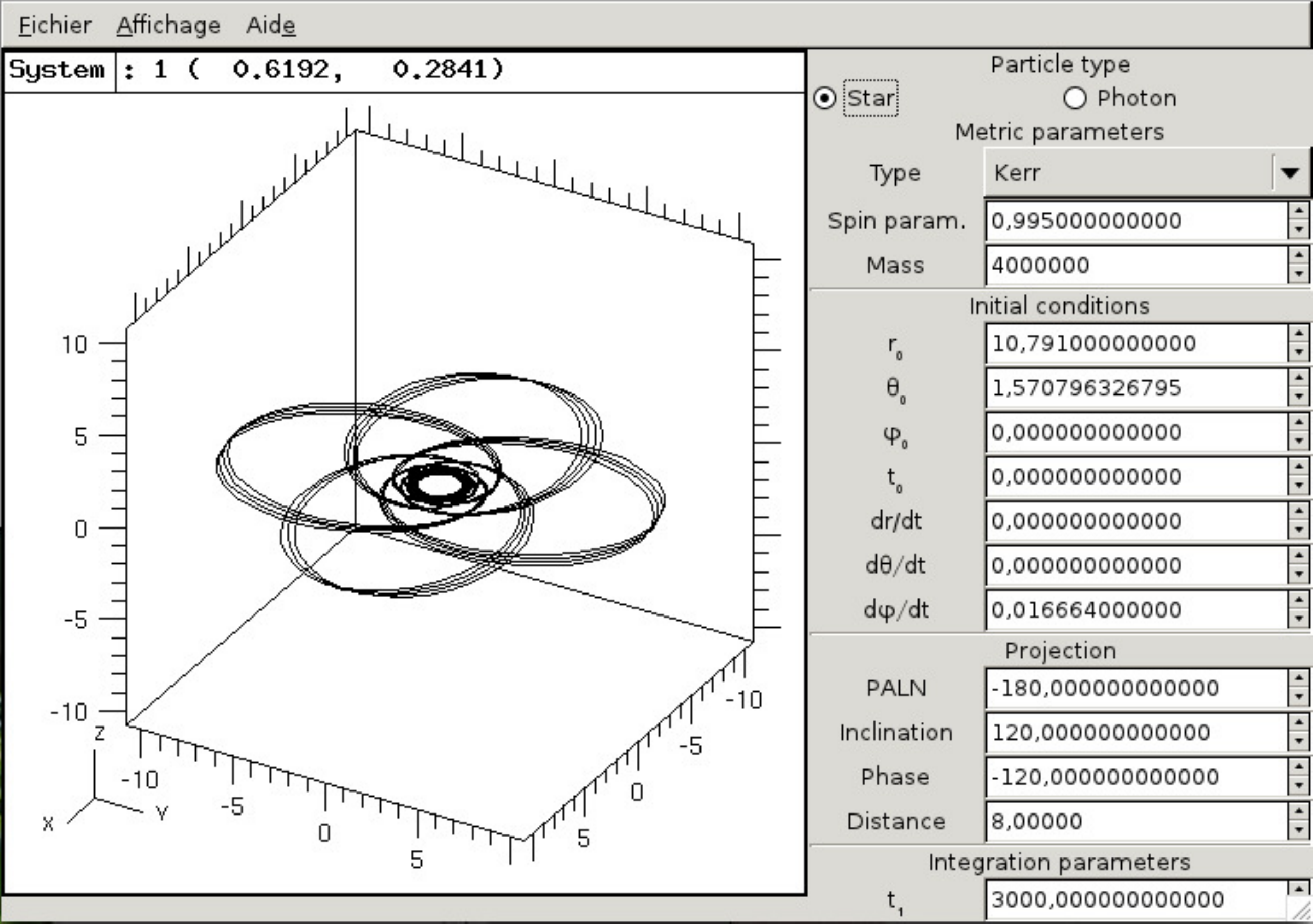}
	\caption{Screen capture of GYOTO's graphical user interface depicting an equatorial orbit in a quasi-extremal Kerr metric. It is possible to interactively change the parameters of the orbit thanks to the tabs on the right column.}
	\label{gyotoy}
\end{figure*}
\item all the core functionnalities of GYOTO are packaged in a C++ shared library (\texttt{libgyoto}) and can therefore be easily accessed from the code.
\end{itemize}

GYOTO has been designed with modularity in mind. A simple plug-in system allows users to easily add new metrics and new astrophysical objects without modifying or even recompiling GYOTO. The standard objects and metrics provided in GYOTO are also implemented in plug-ins, so examples are readily available.

The GYOTO code itself is not (yet) parallel, but the ray-tracing concept is inherently parallelizable. Both the \texttt{gyoto} utility and the \texttt{ygyoto} Yorick extension package are designed to split image computation in regions and can be run from standard job submission systems such as torque, thereby performing an effective parallelization.

GYOTO has been tested on recent versions of Linux and Mac OS X and should run fairly easily on any UNIX-like system.
The code is available online at the following URL: \texttt{http://gyoto.obspm.fr}. A user guide is available there, that gives details about the code architecture, as well as about the software prerequisites needed to run GYOTO.

\section{Conclusion}

We have developed a new ray-tracing code, GYOTO. The code is public, and is designed in order to be easily handled by the end user, even if not specialist of ray-tracing. 
GYOTO is able to integrate null and timelike geodesics in the Kerr metric, as illustrated above for various astrophysical objects surrounding black holes, that are already implemented.
Written in C++, GYOTO is particularly adapted to be developed according to the specific needs of the user, by virtue of its object oriented syntax.

A specificity of GYOTO as compared with other ray-tracing codes is its ability to handle numerical metrics resulting from 3+1 numerical relativity computations.  The development of this aspect of the code will certainly lead in the near future to interesting and new astrophysical results.

The near future of GYOTO will be devoted to the development of new astrophysical objects and to the diversification of the handled numerically computed metrics. The possibility to parallelize the code by using GPU will also be investigated.

\ack
FHV thanks H\'elo\"ise M\'eheut for having provided the numerical data necessary for the computation of the Rossby wave instability image.
This work was supported by grants from R\'egion Ile-de-France and by the ANR grant 06-2-134423 \emph{M\'ethodes math\'ematiques pour la relativit\'e g\'en\'erale}.

\appendix
\section*{Appendix: Convergence test}
\setcounter{section}{1}

\begin{figure*}
\centering
	\includegraphics[width=8cm,height=8cm]{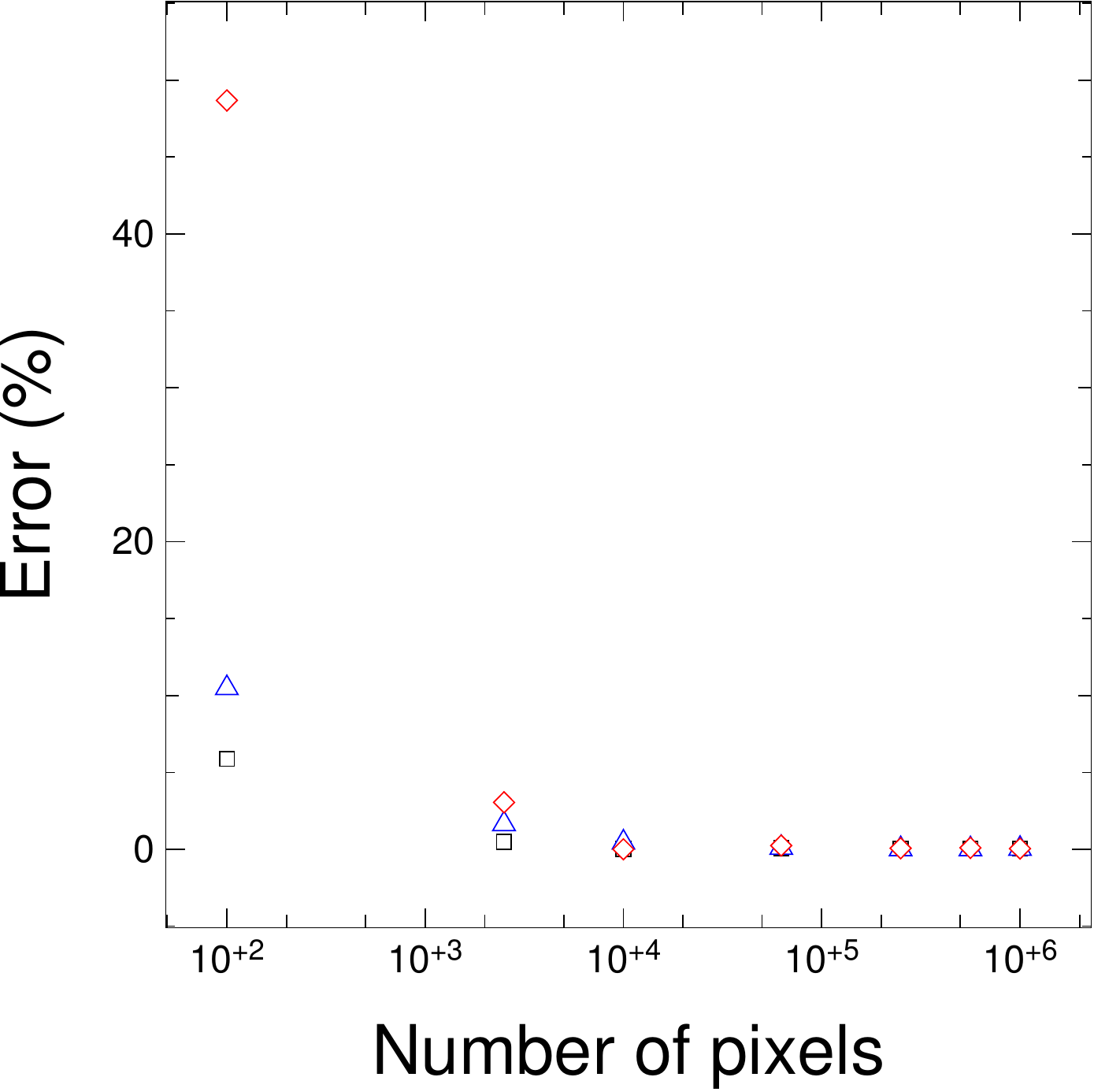}
	\caption{Convergence test for GYOTO. This graph shows the percentage of error for the normalized flux emitted by a thin infinite disk surrounding a black hole of spin 
parameter $a=0$ (black square), $a=0.5M$ (blue triangle) or $a=0.9M$ (red diamond), as a function of the total number of pixels of the GYOTO screen. The percentage of error for $10^{6}$ pixels is less than $0.05\,\%$.}
	\label{cvtest}
\end{figure*}

Figure~\ref{cvtest} shows a convergence test of the GYOTO algorithm. We have considered a thin infinite accretion disk, as described in section~\ref{sec:thindisk}, surrounding a black hole with different values of spin parameter. The observed flux was computed according to equation~(\ref{deffluxGyoto}) for different values of the total number of pixels of the GYOTO screen. The flux reference value is defined as the flux obtained with a screen of $N_{\mathrm{pix}}=2000\times2000$ pixels. The percentage of error, as compared to this reference value, is then determined for different values of $N_{\mathrm{pix}}$.

Figure~\ref{cvtest} shows that GYOTO converges very quickly to very small errors. The error is already below $1\,\%$ for $N_{\mathrm{pix}}=100\times100$. With $N_{\mathrm{pix}}=1000\times1000$, the value used for all the figures shown in this article, the error is less than $0.05\,\%$.

The numerical accuracy of GYOTO is thus very satisfactory. Regarding the speed of GYOTO, the computing time necessary to obtain figure~\ref{diskBL} is around 15~min on a laptop under Mac OS X on one core of a $2.4$ GHz Intel Core 2 Duo processor. The image has $N_{\mathrm{pix}}=1000\times 1000$, the observer being at a distance $r=100\,M$ from the black hole.

\section*{References}
\begin{thebibliography}{22}
\bibitem{cunningham73} Cunningham C T and Bardeen J M 1973 {\it Astrophys. J.} {\bf 183} 237
\bibitem{cunningham75} Cunningham C T 1975 {\it Astrophys. J.} {\bf 202} 788
\bibitem{luminet79} Luminet J-P 1979 {\it Astron. Astrophys.} {\bf 75} 228
\bibitem{fabian89} Fabian A C, Rees M J, Stella L and White N E 1989 {\it Mon. Not. R. Astron. Soc.} {\bf 238} 729
\bibitem{hameury94} Hameury J, Marck J A and Pelat D 1994 {\it Astron. Astrophys.} {\bf 287} 795
\bibitem{fanton97} Fanton C, Calvani M, de Felice F and Cadez A 1997 {\it Publi. of the Astron. Soc. of Japan} {\bf 49} 159
\bibitem{fuerst04} Fuerst S V and Wu K 2004 {\it Astron. Astrophys.} {\bf 424} 733
\bibitem{li05} Li L-X, Zimmerman E R, Narayan R and McClintock J R 2005 {\it Astrophys. J. Suppl.} {\bf 157} 335
\bibitem{wu06} Wu K, Fuerst S V and Lee K 2006 {\it Chinese Journal of Astronomy and Astrophysics Supplement} {\bf 6} 205
\bibitem{dexter09} Dexter J and Agol E 2009 {\it Astrophys. J.} {\bf 696} 1616
\bibitem{karas92} Karas V, Vokrouhlicky D and Polnarev A G 1992 {\it Mon. Not. R. Astron. Soc.} {\bf 259} 569
\bibitem{marck96} Marck J A 1996 {\it Class. Quantum Grav.} {\bf 13} 393
\bibitem{broderick06} Broderick A E and Loeb A 2006 {\it Mon. Not. R. Astron. Soc.} {\bf 367} 905
\bibitem{schnittman06} Schnittman J D, Krolik J H and Hawley J F 2006 {\it Astrophys. J.} {\bf 651} 1031
\bibitem{levin08} Levin J and Perez-Giz G 2008 {\it Phys. Rev.} D {\bf 77} 103005
\bibitem{muller09} M\"uller T 2009 {\it Gen. Relativ. Gravit.} {\bf 41} 541
\bibitem{noble07} Noble S C, Leung P K, Gammie C F and Book L G 2007 {\it Class. Quantum Grav.} {\bf 24} 259
\bibitem{dolence09} Dolence J C, Gammie C F, Moscibrodzka M and Leung P K 2009 {\it Astrophys. J. Suppl.} {\bf 184} 387
\bibitem{birkl07} Birkl R, Aloy M A, Janka H-Th and M\"uller E 2007 {\it Astron. Astrophys.} {\bf 463} 51
\bibitem{harikae10} Harikae S, Kotake K and Takiwaki T 2010 {\it Astrophys. J.}, {\bf 713} 304
\bibitem{broderick04} Broderick A E 2004 {\it Radiative transfer in accreting environments} (PhD Thesis, California Institute of Technology)
\bibitem{zamaninasab10} Zamaninasab {\it et al} 2010 {\it Astron. Astrophys.} {\bf 510} A3
\bibitem{shcherbakov11} Shcherbakov R V and Huang L 2011 {\it Mon. Not. R. Astron. Soc.} {\bf 410} 1052
\bibitem{MulleG10}
M\"uller T and Grave F 2010 {\it Comput. Phys. Com.} {\bf 181} 413
\bibitem{MulleF11}
M\"uller T and Frauendiener J. 2011 {\it Eur. J. Phys.} {\bf 32} 747
\bibitem{psaltis10} Psaltis D and Johannsen T 2010 {\it Preprint} arXiv:astro-ph/1011.4078
\bibitem{mtw73} Misner C W, Thorne K S and Wheeler J A 1973 {\it Gravitation} (San Fransisco: W H Freeman and Company) 
\bibitem{carter68} Carter B 1968 {\it Phys. Rev.} {\bf 174} 1559
\bibitem{numrec} Press W H, Flannery B P, Teukolsky S A and Vetterling W T 1988 \textit{Numerical recipes in C} (Cambridge: Cambridge University Press)
\bibitem{mihalas84} Mihalas D and Mihalas B 1984 {\it Foundations of Radiation Hydrodynamics} (Oxford: Oxford University Press)
\bibitem{page74} Page D N and Thorne K S 1974 {\it Astrophys. J.} {\bf 191} 499
\bibitem{meheut10} M\'eheut H, Casse F, Varni\`ere P and Tagger M 2010 {\it Astron. Astrophys.} {\bf 516} A31
\bibitem{toth96} T\'oth G 1996 {\it Astrophys. Lett. Comm.} {\bf 34} 245
\bibitem{falanga07} Falanga M, Melia F, Tagger M, Goldwurm A and B\'elanger G 2007 {\it Astrophys. J.} {\bf 662} L15
\bibitem{abramowicz78} Abramowicz M, Jaroszynski M and Sikora M 1978 {\it Astron. Astrophys} {\bf 63} 221
\bibitem{abramowicz08} Abramowicz M 2009 {\it ASP Conference Series} {\bf 403} 29
\bibitem{narayan95} Narayan R and Yi I 1995 {\it Astrophys. J.} {\bf 452} 710
\bibitem{straub11} Straub O, Vincent F H, Abramowicz M, Gourgoulhon E and Paumard T 2011 {\it in prep.}
\bibitem{fuerst07} Fuerst S V and Wu K 2007 {\it Astron. Astrophys.} {\bf 474} 55
\bibitem{gourgoulhon06} Gourgoulhon E 2007 3+1 Formalism and Basis of Numerical Relativity {\it Preprint} arXiv:gr-qc/0703035
\bibitem{Alcub08}
Alcubierre M 2008 {\it Introduction to 3+1 Numerical Relativity}
(Oxford: Oxford University Press) 
\bibitem{BaumgS10}
Baumgarte T W and Shapiro S L 2010 {\it Numerical Relativity.
Solving Einstein's Equations on the Computer}
(Cambridge: Cambridge University Press). 
\bibitem{vincent11} Vincent F H, Gourgoulhon E and Novak J 2011 
Geodesic equation within the 3+1 formalism,
{\it in prep.}
\bibitem{GrandN09}
Grandcl\'ement P and Novak J 2009 :
{\it Living Rev. Relat.} {\bf 12}, 1 (2009);
\texttt{http://www.livingreviews.org/lrr-2009-1}
\bibitem{lorene}
\texttt{http://www.lorene.obspm.fr}
\bibitem{gourgoulhon10} Gourgoulhon E 2010 An introduction to the theory of rotating relativistic stars {\it Preprint} arXiv:1003-5015
\end{thebibliography}
\end{document}